\documentclass[pdflatex,sn-aps]{sn-jnl}
\usepackage{amsmath}
\usepackage{amssymb}
\usepackage{graphicx}

\jyear{2022}%

\begin{document}

    \title[Data-driven estimation of transfer integrals in undoped cuprates]{Data-driven estimation of transfer integrals in undoped cuprates}


    \author*[1]{\fnm{Denys~Y.}~\sur{Kononenko}}\email{d.kononenko@ifw-dresden.de}
    \author[1]{\fnm{Ulrich~K.}~\sur{R{\"o}{\ss}ler}}
    \author[1,2]{\fnm{Jeroen}~\spfx{van~den}~\sur{Brink}}
    \author*[1]{\fnm{Oleg}~\sur{Janson}}\email{olegjanson@gmail.com}

    \affil[1]{\orgname{Institute for Theoretical Solid State Physics}, \orgaddress{\city{Dresden}, \postcode{01069}, \country{Germany}}}
    \affil[2]{\orgdiv{Institute for Theoretical Physics}, \orgname{TU Dresden}, \orgaddress{\city{Dresden}, \postcode{01069}, \country{Germany}}}

    \abstract{Undoped cuprates are an abundant class of magnetic insulators, in which the synergy of rich chemistry and sizable quantum fluctuations leads to a variety of magnetic behaviors. Understanding the magnetism of these materials is impossible without the knowledge of the underlying spin model. The typically dominant antiferromagnetic superexchanges can be accurately estimated from the respective electronic transfer integrals. Density functional theory calculations mapped onto an effective one-orbital model in the Wannier basis are an accurate, albeit computationally cumbersome method to estimate such transfer integrals in cuprates. We demonstrate that instead an Artificial Neural Network (ANN), trained on the results of high-throughput calculations, can predict the transfer integrals using the crystal structure as the only input. Descriptors of the ANN model encode the spatial configuration and the chemical composition of the local crystalline environment. A virtual toolbox employing our model can be readily employed to determine leading superexchange paths as well as for rapidly assessing the relevant spin model in yet unknown cuprates.}




    \maketitle

    \section{Introduction}\label{sec:intro}
    The data-driven approach accompanied by modern machine-learning (ML) techniques
    becomes an increasingly important tool of scientific investigations across many
    domains of physics. From quantum to fluid mechanics~\cite{tsubaki.prl.20,
    brunton.annurev-fluid.20} learning from data facilitates descriptions of
    complex phenomena for which analytical approaches are prohibitively challenging.
    The adoption of ML in solid-state physics and material science is
    particularly appealing: the sheer amount of collected experimental and
    computed records propels the community to design data-driven frameworks for prediction
    of materials properties~\cite{scmidt.npj.compmat.19, xie.prl.20}.

    The application of ML for problems of solid-state physics is not
    straightforward. One of the key challenges is to represent periodic (crystalline)
    or finite (molecule, local crystal environment, etc.) atomic systems as
    descriptors -- data structures amenable to ML methods. Such descriptors must be
    invariant with respect to the choice of the unit cell (crystals) or to global
    rotations in a finite system.  Several classes of descriptors have been
    developed for material properties prediction: Coulomb
    matrix~\cite{rupp.prl.12}, partial radial distribution
    function~(PRDF)~\cite{schtt.prb.2014}, smooth overlap of atomic
    positions~(SOAP)~\cite{bartok.prb.2013}, diffraction
    fingerprint~(DF)~\cite{ziletti.natcomm.2018} and three-dimensional~(3D) Zernike
    descriptor (3DZRD)~\cite{venkatraman.joc.09}.  The latter were designed for the
    characterization of 3D shapes~\cite{novotni2003, novotni.shape.04} and
    successfully employed for comparison of molecules~\cite{sael.prot.2008,
    mak.jmgm.2008, sael2008}. These descriptors are invariant with respect to the
    number of chemical species in the dataset, they store detailed information
    about the spatial arrangement, and do not require additional simulation
    software. These features as well as the compact size of the resulting data
    structures make 3DZRD ideally suited for a description of diverse, dissimilar
    crystalline environments.

    Another key challenge is the construction of descriptors that represent
    material properties. For instance, it is widely accepted that electronic,
    magnetic, and topological properties of bulk materials are rooted -- in a
    highly nontrivial way --- in their electronic structure. However, using the
    complete band structure of a material as a universal descriptor is not possible
    for a number of reasons: different number of bands, non-universal
    discretization of the Brillouin zone, huge dimensionality etc. A more practical
    approach is to restrict the description to the states relevant for the physical
    quantity of interest. Naturally, this is possible only for a certain class of
    materials and only for specific physical property.

Following this idea, we apply a data-driven approach to assess spin models in
undoped cuprates -- stoichiometric inorganic materials containing divalent
copper and oxygen atoms. In contrast to their doped counterparts -- the
high-temperature cuprate superconductors~\cite{Plakida2010htc} -- undoped
cuprates are magnetic insulators with the $3d^9$ electronic configuration of
Cu$^{2+}$. The sizable Jahn-Teller distortion lifts the orbital degeneracy,
giving rise to half filling and localized $S$=$\frac12$ spins. Owing to the
plethora of structure types and the quantum limit assured by $S$=$\frac12$,
undoped cuprates exhibit a variety of magnetic
behaviors~\cite{sahadasgupta2021}, from simple quantum dimers and spin chains
-- to exotic collective behaviors such as the spin-liquid regime in
herbertsmithtite $\gamma$-Cu$_3$Zn(OH)$_6$Cl$_2$~\cite{helton07}, Bose-Einstein
condensation of magnons in Han purple BaCuSi$_2$O$_6$~\cite{jaime04}, or bound
magnon states in volborthite
Cu$_3$V$_2$O$_7$(OH)$_2\cdot2$H$_2$O~\cite{kohama19}.

Understanding the magnetic properties of cuprates requires the knowledge of the
underlying spin model. While exchange anisotropies are generally present and
can alter the magnetic properties, the backbone of spin models are isotropic
interactions, and the relevant Heisenberg Hamiltonian is the following sum:
       \begin{align}\label{eq:heisenberg}
           \mathcal{H} = \frac12\sum_{ij} J_{ij}\left(\mathbf{S}_i \cdot \mathbf{S}_j\right),
       \end{align}
where $\mathbf{S}_i$ and $\mathbf{S}_j$ are spin operators on sites $i$ and
$j$. The set of relevant magnetic exchange integrals $\{J_{ij}\}$ determines
the spin model. It is important to note that each individual $J_{ij}$ term is a
sum of antiferromagnetic ($J^\text{AF}_{ij} < 0$) and ferromagnetic
($J^\text{FM}_{ij} < 0$) contributions that are driven by competing
processes~\cite{Goodenough_Magnetism}. Commonly, $J^\text{AF}_{ij} \gg
\lvert{}J^\text{FM}_{ij}\rvert$, with the exception of short-range exchanges
for which the ferromagnetic contribution can become dominant.

The antiferromagnetic contribution is a textbook example of the
superexchange mechanism and can be derived via second-order perturbation theory
of the Hubbard model in the strong-coupling limit at half-filling as
$J^\textsc{AF}_{ij} = 4 t_{ij}^2 /U_\text{eff}$~\cite{hubbard63, anderson59,
auerbach2000}. Here, $U_\text{eff}$ is the Coulomb repulsion within an
effective molecularlike orbital, which in most cuprates is dominated by
$3d_{x^2 - y^2}$ orbital of Cu and $\sigma$-bonded $2p$ orbitals of O. There is
empirical evidence that $U_\text{eff}$ from the range 4--5 eV gives a proper
description of the magnetism of cuprates~\cite{belik2004, johannes2006,
janson2009}. Hence, the knowledge of transfer integrals $t_{ij}$ paves the way
to a quantitative assessment of the spin model in the majority of cuprate
materials. Yet, extracting $t_{ij}$ directly from the structural information is
essentially impossible; instead, it requires first-principles calculations
followed by an additional modeling.

To overcome this challenge, we propose a data-driven approach for prediction of
transfer integrals in cuprates, which requires the crystal structure as the
only input.  Our approach is based on the local crystal environment description
utilizing 3DZRD.  The local crystal environment descriptor is used as input for
the ML model which is trained on the results of high-throughput
density-functional-theory (DFT) calculations for hundreds of cuprate materials.
DFT calculation for each material is followed by automatized Wannierization and
a manual quality control. The trained model is wrapped into a freely accessible
web application \footnote{https://smc-t.ifw-dresden.de/} that can be used for a
quick estimation of relevant transfer paths in new cuprate materials.

The paper is organized as follows: Section 2 describes the high-throughput DFT 
calculations and the dataset of transfer integrals. Section 3 details the descriptor 
for the Cu..Cu bonds. In Section 4 we compare three 
different ML approaches for predicting transfer 
integrals and estimate the accuracy by a cross-validation procedure (CV). In the last section, we discuss the performance of our ML model for different classes of cuprates. In particular, for the parent (undoped) compounds of high-temperature 
superconducting cuprates we show that the ANN model quantitatively captures the ratio 
between nearest and next-nearest-neighbor transfer integrals. 

    \section{Dataset generation}
    We start with the description of the high-throughput DFT calculations employed
    for the generation of the dataset of transfer integrals. The list of materials
    contains 672 unique structures of undoped cuprates. The structures were
    filtered out from the 10\,710 cuprate structures stored in the Inorganic
    Crystal Structure Database (ICSD)~\cite{bergerhoff.crystallographic.1987}. For this screening, the
    following criteria were consecutively applied: (i) the presence of Cu$^{2+}$
    ions, (ii) electroneutrality (zero total charge), (iii) absence of sites with
    fractional occupancies, (iv) the minimal inter-atomic distance of 0.5\,\r{A},
    and (v) the absence of other magnetic atoms beyond Cu~\cite{suppl}. The latter criterion is necessary to
    filter out compounds with multiple magnetic atoms, where the presence of
    additional bands in the relevant energy range may render the effective
    one-orbital model inapplicable and its results misleading.  For the
    analysis of the crystal structures we used the pymatgen library~\cite{Ong2013pymatgen} for Python.

    For each structure, we performed DFT calculations to construct Wannier
    Hamiltonians and subsequently determined the transfer integrals. All DFT
    calculations were performed using the generalized gradient approximation
    (GGA)~\cite{perdew.prl.1996} with the full potential code FPLO of version
    18.00-52~\cite{koepernik.prb.1999}. The computational workflow comprised
    several steps. First, scalar-relativistic nonmagnetic DFT calculations were
    carried out and the Hellmann-Feynman forces were calculated. 
    Second, for hydrogen-containing compounds whose calculated forces exceeded
the threshold of 0.1\,eV/\r{A}, we optimized the internal coordinates of H
atoms within GGA. The rationale behind this step are largely inaccurate H
positions as determined by x-ray diffraction (which is by far most common
method of structure determination). Since a considerable number of cuprates
contain hydrogen, typically as hydroxyl groups or water molecules, inclusion of
such partly optimized structures allowed us to considerably extend the data
set. All other cuprates whose forces exceeded the threshold 0.1\,eV/\r{A} were
discarded. Third, we calculated the orbital-resolved density of states (DOS)
and band 
    structure.
    From the orbital-resolved DOS, the energy interval which contains the copper 
    $3d_{x^2 - y^2}$ bands was
    determined. The energy interval is selected such that the contribution of the 
    magnetic $3d_{x^2 - y^2}$ orbital in the total density of Cu $3d$ states exceeds 5 
    \%.
    The determined energy window $[\mathscr{E}_\textsc{min}, \mathscr{E}_\textsc{max}]$
    was adopted for Wannierization in the next step.
    Fourth, the Wannier transformation procedure was performed to obtain the
    effective one-orbital Hamiltonian $H$ in the Wannier basis. We used copper
    $3d_{x^2-y^2}$ orbitals as projectors and the interval $[\mathscr{E}_\textsc{min},
    \mathscr{E}_\textsc{max}]$ as the energy window to construct the Wannier
    functions~(WF). The latter is necessary to discriminate the target antibonding
    orbital (crossing the Fermi energy) from its bonding sibling at the
    bottom of the valence band. The transfer integral between two WF $w_i$ and $w_j$
    placed at copper sites $i$, $j$ is determined as (real) Hamiltonian matrix
    element $t_{ij} = \langle{}w_i\lvert{}H\rvert{}w_j\rangle$. The details on the 
    construction of Wannier functions and the construction of the respective
tight-binding models are provided in the papers~\cite{koepernik2021arxiv,
eschrig2009}.

\begin{center}
            \begin{figure}
                \includegraphics[width=.667\textwidth]{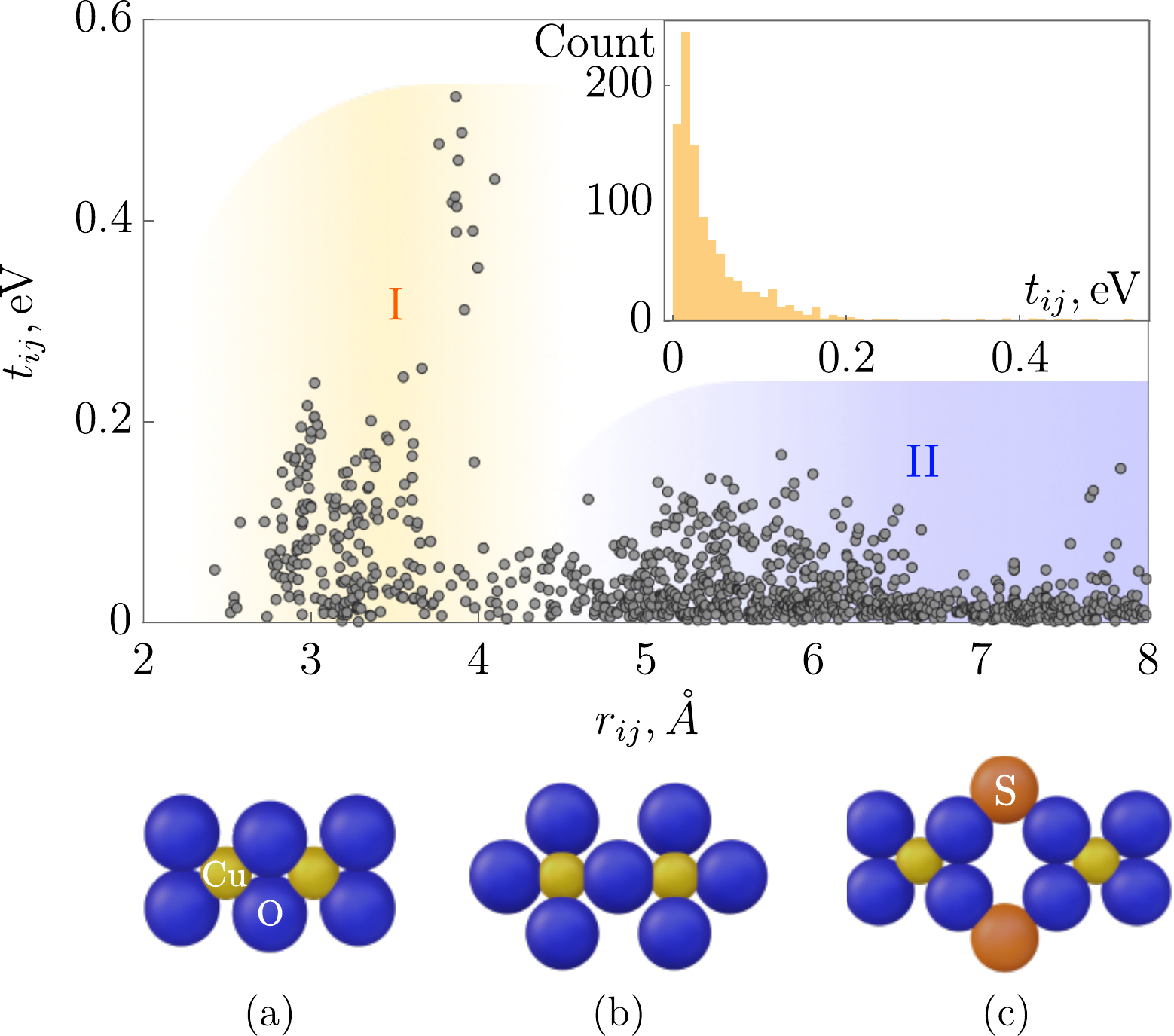}
                \caption{(Color online) Transfer integrals obtained from the DFT calculations  as a function of the Cu..Cu distance. Region I harbors transfer integrals between edge-sharing (a) and corner-sharing (b) CuO$_4$ plaquettes, while region II is dominated by transfer integrals between CuO$_4$ plaquettes that do not share oxygen atoms (c). The inset shows the distribution of the computed transfer integrals.
                }
                \label{fig:fig_data}
                \includegraphics[width=.667\textwidth]{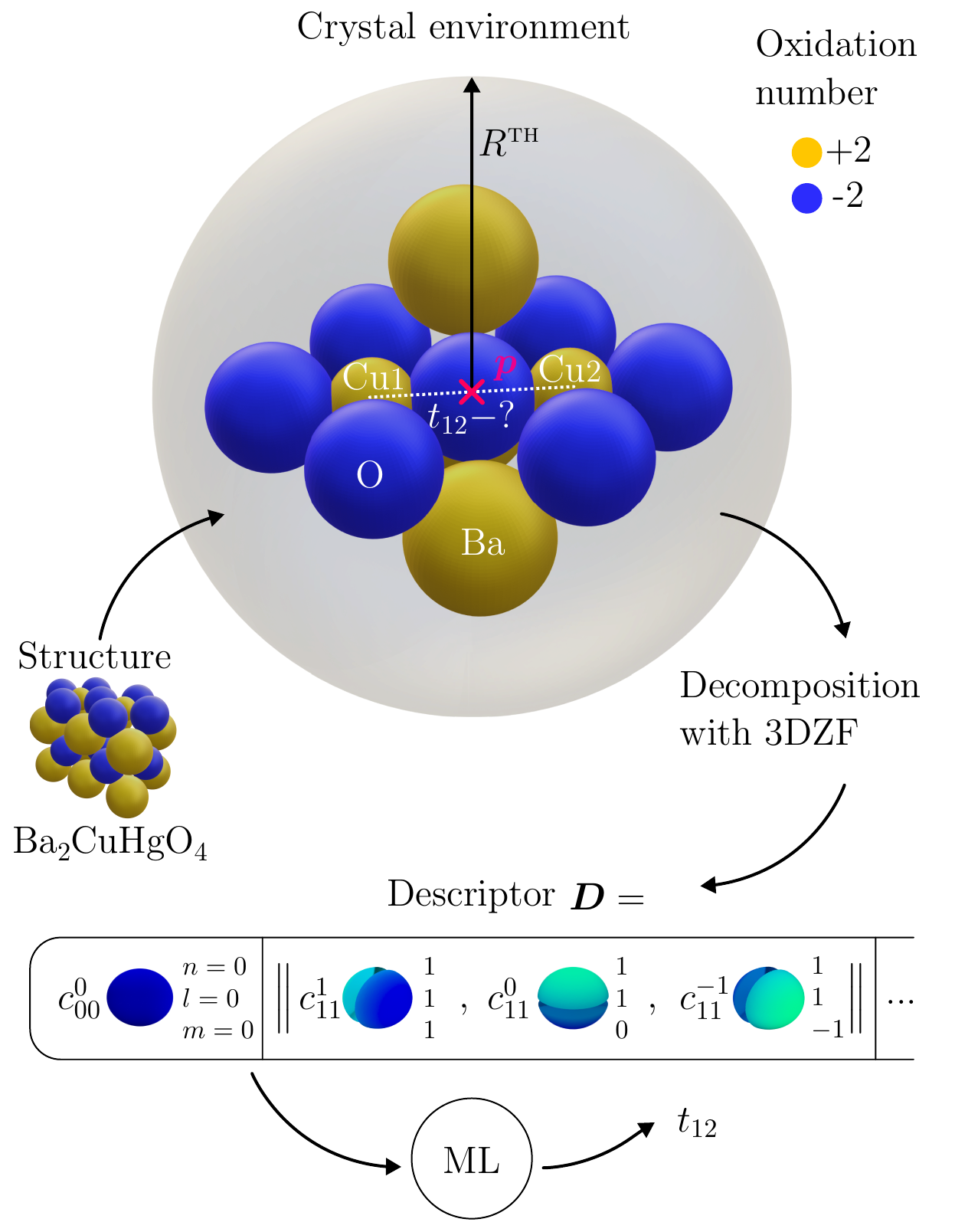}
                \caption{(Color online) Schematics of the workflow: selection of the local crystal environment from the cuprate crystal structure, generation of rotationally invariant descriptor $\vec{D}$ via decomposition of the local crystal environment function in the truncated basis of 3DZF and prediction of the transfer integral $t_{ij}$ with ML algorithm trained on the dataset from DFT calculations. The illustrating example Ba$_2$CuHgO$_4$ (ICSD Identifier 75724) hosts pairs of corner-sharing CuO$_4$ square-like plaquettes.
                }
                \label{fig:fig_crenv}
            \end{figure}
        \end{center}

    After the calculation pipeline was completed, we obtained a list of
transfer integrals $\{t_{i j}\}$ that connected $i$-th and $j$-th copper sites
situated at the distance $r_{ij}$ from each other for all valid
structures~\cite{suppl}.  For construction of the dataset we select transfer
integrals larger than 5 meV with Cu..Cu spacing less than 8 \r{A}.
    The distribution of calculated transfer integrals $t_{ij}$ among Cu..Cu distances
    is shown in Fig.~\ref{fig:fig_data}. The crystal chemistry of cuprates sets a
    natural lower limit for the bond lengths; accordingly, there is no transfer
    integral with the distance less than 2.4~\r{A}\ in the dataset. Remarkably,
    for the vast majority of transfer integrals, the absolute values are below
    0.2\,eV. This natural imbalance of the dataset will inevitably affect the
    performance of predictions.

        \section{Crystal Environment Descriptor}
        To describe the crystal environment, we first determine the midpoint $\vec{p}$  between a given pair of copper atoms and build a sphere with the empirically determined threshold radius $R^\textsc{th} = \max(4, r_{i j} / 2 + 0.2)$ \r{A} centered at $\vec{p}$. Next, all atoms in the sphere are enlisted in the crystal environment alongside with nearest neighbors of $i$ and $j$. We consider nearest neighbors as atoms distanced from $i$ or $j$ not farther than 2.5 \r{A}. After the local crystal environment is assembled, we shift the coordinate system origin to the centroid (the point between copper pair) $\vec{p}$ and normalize atoms coordinates by $r_0 = 6$ \r{A} to fit the crystal environment in the unit ball.
        To construct a robust representation of the local crystal environment we introduce the piecewise function of site positions $\mathcal{I}(\vec{r})$. The function $\mathcal{I}$ equal to the $q$-th atom oxidation number $O_q$ in the ball with center at the position of  $q$-th atom $\vec{r}_q$ and radius $R_q$ equals to the ionic radius of the atom
        \begin{align}\label{eq:cryst_env_func}
        \mathcal{I}(\vec{r}) =
        \begin{cases}
        O_q & \Vert \vec{r}_q - \vec{r}\Vert \leq R_q, \\
        0 & \text{otherwise}.
        \end{cases}
        \end{align}
        The normalization factor $r_0$ is a sum of the maximal considered Cu-Cu distance $\max \Vert \vec{r}_{i j} \Vert = 4$ \r{A} and a maximal considered ionic radius $2$ \r{A}.
        The function $\mathcal{I}(x,y,z)$ describes the spatial configuration and chemical composition of the crystal environment placed in the unit ball with $x^2 + y^2 + z^2 \leq 1$. An example of the local crystal environment defined by~\eqref{eq:cryst_env_func} is shown in Fig.~\ref{fig:fig_crenv}.
     
        We describe the selected crystal environment $\mathcal{I}$ in the form of a finite-dimensional vector.
        Such representation provides a robust way for numerical operations with crystal environments, e.g. similarity and sorting.
        To obtain the finite vector representation of the crystal environment we decompose the $\mathcal{I}(x,y,z)$ in the truncated basis of three-dimensional (3D) Zernike functions (3DZF)  $Z^{m}_{nl}$ which are defined as follows~\cite{canterakis96, canterakis99, novotni.shape.04}
        \begin{align}\label{eq:zernike_func_3d}
        \begin{split}
        Z_{nl}^{m}(r, \theta, \phi) &= R_{nl}(r) Y_{lm}(\theta, \phi), \\
        R_{nl}(r) &= \sum_{\nu = 0}^{(n-l)/2} Q_{l\nu} r^{2 \nu + l}, \\
        Q_{l\nu} &= \dfrac{(-1)^{k + \nu}}{4^{k}} \sqrt{\dfrac{2l + 4 k + 3}{3}} \dfrac{ \binom{2 k}{k} \binom{k}{\nu} \binom{2 (k + l + \nu)+1}{2 k} }{\binom{k + l + \nu}{k}},
        \end{split}
        \end{align}
        where indices $n$ and $l$ are positive integers which satisfy condition $n \geq l$; $m$ changes from  $-l$ to $l$ with constraint $(n-l)$ is even number; $k = (n - l) / 2$ and $Y_{lm}(\theta, \phi)$ are spherical harmonics, and $(r, \theta, \phi)$ are spherical coordinates~\cite{nist.math.handbook}.
        For convenience, we use Cartesian coordinates $(x,y,z)$ representation
        of 3DZF implying change of coordinates: $\mathcal{Z}_{nl}^{m}(x, y, z) = Z_{nl}^{m}\left(\sqrt{x^2 + y^2 + z^2}, \arctan{\sqrt{x^2 + y^2} / z}, \arctan{y / x}\right)$.
        3DZF form the complete basis of orthogonal functions in the unit ball, so that the function $\mathcal{I}(x,y,z)$ defined in the unit ball $x^2 + y^2 + z^2 \leq 1$ can be expanded in the introduced basis~\cite{morais.mathematics-of-computation.2014}.

        The decomposition coefficients read
        \begin{align}\label{eq:zernike_moment_3d}
        c_{nl}^{m} = \dfrac{1}{V} \int_{-1}^{1}\int_{-1}^{1}\int_{-1}^{1} \mathcal{I}(x, y, z) \mathcal{Z}_{nl}^{m}\left(x,y,z\right) \mathrm{d}x\mathrm{d}y \mathrm{d}z,
        \end{align}
    where the normalization factor is the volume of the unit ball $V = 4 \pi / 3$~\cite{suppl}.

        Note, $c_{nl}^{m}$ is not invariant with respect to rotations of the crystal environment $\mathcal{I}$. Rotationally invariant characteristics can be obtained by assembling the vector $\vec{C}_{nl}$ whose components are all $(2 l + 1)$ coefficients with different $m$ for given pair of $n$ and $l$. The norm of obtained vector $\Vert\vec{C}_{nl}\Vert = C_{nl}$ determined as
        \begin{align}\label{eq:zernike_rot_inv_moment_3d}
        C_{nl} = \left\Vert c_{nl}^{0}, ..., c_{nl}^{l-1}, c_{nl}^{l} \right\Vert,
        \end{align}
        is invariant with respect to the rotation of the crystal environment, thus the pre-alignment is not required.

        We introduce the finite dimensional vector-descriptor of the crystal environment $\mathcal{I}$ as
        \begin{align}\label{eq:zernike3d_descriptor}
        \vec{D} = \left( C_{00}, C_{11}, C_{20}, ..., C_{n_\textsc{max}l_\textsc{max}}, r_{ij} \right),
        \end{align}
    where copper-copper distance $r_{ij}$ is incorporated into the descriptor as well.
    The size of the descriptor $\vec{D}$ is determined by the cut-off order of the Zernike 3D moments $n_\textsc{max}$ and corresponding $l_\textsc{max}$ in the truncated basis. The size of the 3DZF basis grows with $n_\textsc{max}$
    as the sum of the series $\sum_{n=0}^{n_\textsc{max}}(n^2 + 3 n + 2) / 2$. In the present work, we chose the cut-off order $n_\textsc{max} = 25$.
    The vector  $\vec{D}$ encodes the information about spatial configuration and chemical composition of the crystal environment, allowing the introduction of the mapping of $\vec{D}$ on the transfer integral.

        \begin{center}
                \begin{figure}[!t]
                        \includegraphics[width=.667\textwidth]{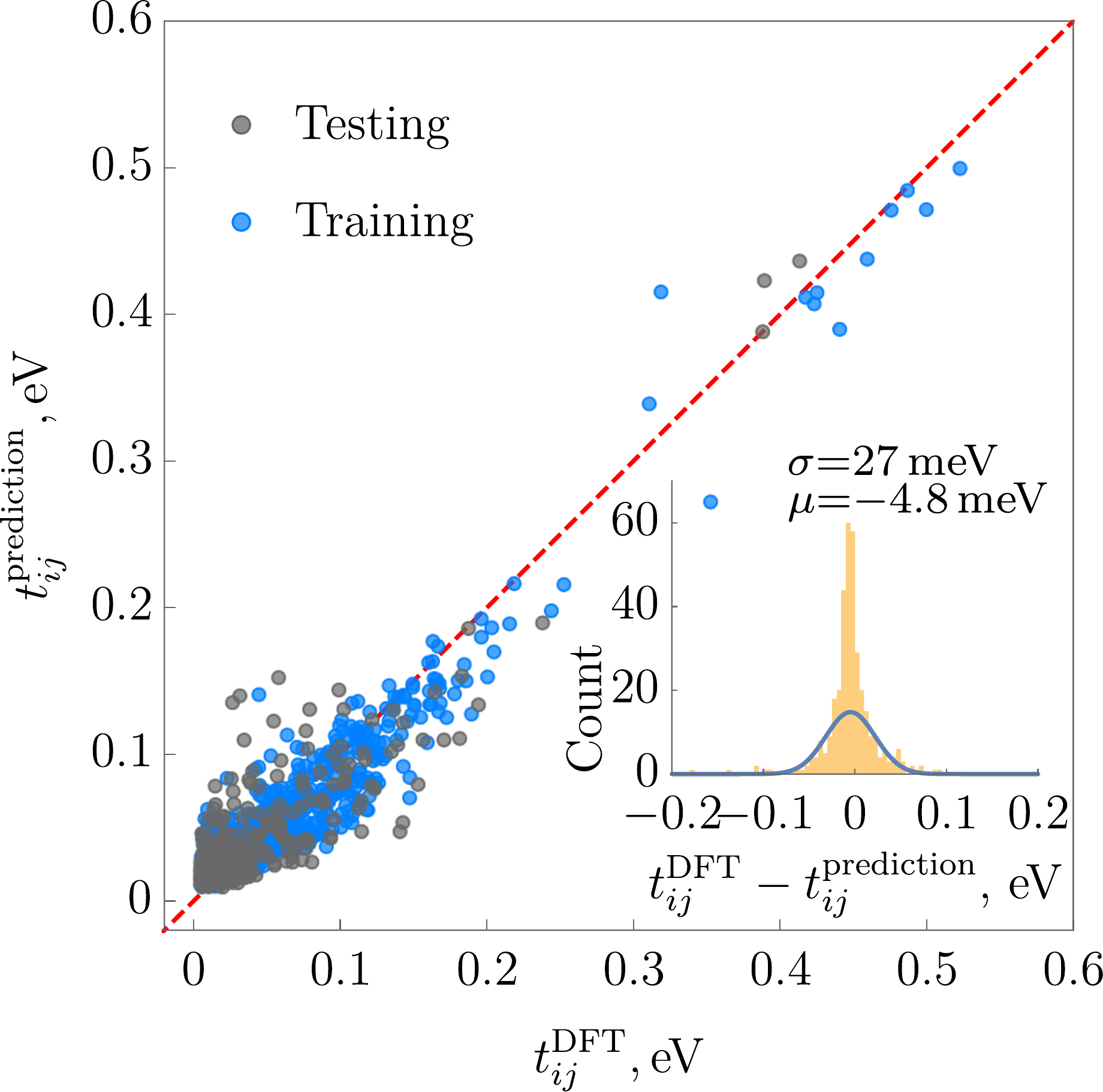}
                        \caption{(Color online) Performance of the ensemble ANN model on the testing and training datasets for random split.
                        The ensemble ANN shows $R^2 = 0.7$, RMSE = 28 meV and MAE = 18 meV on the testing dataset and $R^2 = 0.9$, RMSE = 19 meV and MAE = 11 meV on the training dataset.
                        The inset figure shows the distribution of the ensemble ANN model error for the test dataset with $\mu$ and $\sigma$ are mean value and standard deviation of the errors. The solid line corresponds to the normal distribution with parameters $\mu$ and $\sigma$.
                        }
                        \label{fig:model_eval}
                \end{figure}
        \end{center}

        \section{Transfer Integral Prediction}
        Our high-throughput DFT calculations yielded $N = 1800$ 
        local crystal environments $\{\vec{D}\}$ with corresponding transfer integrals $\{t_{ij}\}$. 
        We build the ML model to predict the continuous-valued attribute $t_{ij}$ associated with the local crystal environment descriptor $\vec{D}$.
        To solve this regression problem, we tested the following models: (i) linear (LIR), (ii) random forest~(RFR)~\cite{Breiman_2001} regression models, and (iii) ANN. 
        To achieve robust generalization and stability of the models, we employ the bagging (bootstrap aggregating) ensemble technique~\cite{breiman1996bagging}.
        The main idea behind bagging is to train multiple instances of the same model on different subsets of the training data and then combine their predictions to make the final estimation. For each model in the ensemble, a random sample is drawn with a replacement from the original training dataset. Thus, some data points may appear multiple times in the sample, while others may be left out.
        When making predictions, the individual predictions from each model are combined using a voting method (for classification tasks) or averaging (for regression tasks). In the work, we employ ensemble models with 100 estimators.

        As a metric for the ensemble regression model performance with predictions $\tau$ we use:
        (i) the coefficient of determination
        \begin{align}\label{eq:r2}
        R^2 = 1 - \dfrac{S_\textsc{reg}}{S_\textsc{tot}},
        \end{align}
        where $S_\textsc{reg} = \sum_{p=1}^M (t^p_{ij} - \tau^p)^2$ is a sum of squared residuals of the regression model and $S_\textsc{tot} = \sum_{p=1}^M (t^p_{ij} - \overline{t_{ij}})^2$ is a total sum of squares with $\overline{t_{ij}}$ being the mean value of transfer integral in the test dataset with $M$ samples.

        (ii) the root mean squared error (RMSE)
        \begin{align}\label{eq:rmse}
        \text{RMSE} = \sqrt{\dfrac{1}{M} \sum_{p=1}^M (t^p_{ij} - \tau^p)^2},
        \end{align}
    and (iii) mean absolute error
    \begin{equation}\label{eq:mae}
        \text{MAE} = \dfrac{1}{M} \sum_{p=1}^M \vert t^p_{ij} - \tau^p \vert.
    \end{equation}
    For RMSE, squared errors of the model are included in the average, making this measure more sensitive to outliers. As more variance in predictions, a larger RMSE. The MAE provides a mean of linear scores with all errors weighted equally.

    \begin{table}[h]
        \centering
        \begin{tabular}{l|r|r|r}
                Model & $\overline{R^2}$ & $\overline{\text{RMSE}}$, \text{meV} & $\overline{\text{MAE}}$, \text{meV} \\
                \hline
                LIR      & 0.27 $\pm$ 0.18 & 44 $\pm$ 5 & 28 $\pm$ 1  \\
                RFR      & 0.59 $\pm$ 0.01 & 34 $\pm$ 3 & 21 $\pm$ 1  \\
                ANN      & 0.69 $\pm$ 0.05 & 29 $\pm$ 3 & 18 $\pm$ 1  \\
        \end{tabular}
        \vspace{1em}
        \caption{Results of shuffle-split CV of the selected ensemble models. The average value of $R^2$, RMSE, and MAE on six splits is given alongside the standard deviation.}
        \label{tab:ss_cv_res}
    \end{table}

    \begin{table}[h]
        \centering
        \begin{tabular}{l|r|r|r}
                Model & $\overline{R^2}$ & $\overline{\text{RMSE}}$, \text{meV} & $\overline{\text{MAE}}$, \text{meV} \\
                \hline
                LIR      & 0.27 $\pm$ 0.20 & 43 $\pm$ 6 & 28 $\pm$ 1  \\
                RFR      & 0.57 $\pm$ 0.06 & 34 $\pm$ 5 & 21 $\pm$ 1  \\
                ANN      & 0.69 $\pm$ 0.10 & 28 $\pm$ 3 & 18 $\pm$ 1  \\
        \end{tabular}
        \vspace{1em}
        \caption{Results of $k$-fold CV of the selected ensemble models with six folds. The average value of $R^2$, RMSE, and MAE is given alongside the standard deviation.}
        \label{tab:kf_cv_res}
    \end{table}
    
    For model selection, we employ two CV strategies: the shuffle-split and $k$-fold. In the shuffle-split, the dataset is randomly shuffled and then split into training and test subsets containing a specific percentage of the original data. The procedure is repeated the specified number of iterations.
    In each iteration the model is trained and evaluated accordingly.
    The shuffle-split procedure was implemented using the scikit-learn library~\cite{scikit-learn} with six splits, a test size of 20 \%, and a training size of 80 \% of the entire dataset.
    The results of the CV are presented in the Table~\ref{tab:ss_cv_res}.
    The model selection procedure shows that ensemble ANN has the best performance among the selected models. In particular, ensemble ANN has the lowest average errors, $\overline{\text{MAE}}$ = 18 meV and $\overline{\text{RMSE}}$ = 29 meV  with the standard deviation of 1 and 3 meV respectively.

    In the $k$-fold CV, the entire dataset is split into $k$ approximately equal parts (folds). Each ML model is trained on the $k-1$ folds and evaluated on one fold. The $k$-fold procedure was implemented using the scikit-learn library~\cite{scikit-learn} with $k = 6$. The results of the $k$-fold CV are presented in Table~\ref{tab:kf_cv_res}. Similarly to the shuffle-split CV, $k$-fold CV shows that the ensemble ANN has the best scores with $\overline{\text{MAE}}$ = 18 meV and $\overline{\text{RMSE}}$ = 28 meV with the standard deviation of 1 and 3 meV respectively.

    We also evaluated the ANN model on the random test-train split with 20~\% of the data allocated for the test subset.
    The prediction of transfer integrals for the test set is shown in Fig.~\ref{fig:model_eval} as a scatter plot of the calculated values versus predicted ones.

\section{Discussion}
In this section, we will apply our ensemble ANN model to different classes of cuprates
and discuss its predictive power.  The first example are parent compounds of
high-temperature superconductors (HTSC).  The common structural feature of
these antiferromagnets are cuprate planes formed by corner-sharing CuO$_4$
plaquettes. The Cu-O-Cu angle amounts to 180$^{\circ}$, maximizing electron
transfer between the nearest neighbors ($t_1$) and giving rise to a sizable
antiferromagnetic exchange of about 1500\,K~\cite{coldea01}. In addition, the
favorable mutual orientation of plaquettes boosts the coupling between second
neighbors ($t_2$), as confirmed experimentally~\cite{tanaka04}.  Hence, the
magnetic properties of undoped HTSC cuprates are described by the frustrated
square-lattice model, with competing first- and second-neighbor
antiferromagnetic exchanges.  Interestingly, ramifications of this competition
go far beyond the magnetism: the $t_2/t_1$ ratio shows correlations with the
superconducting transition temperature~\cite{Pavarini2001}. Thus, an accurate
estimation of this ratio is crucial for understanding the physics of HTSC
materials.

\begin{center}
        \begin{figure}[!t]
        \includegraphics[width=0.9\textwidth]{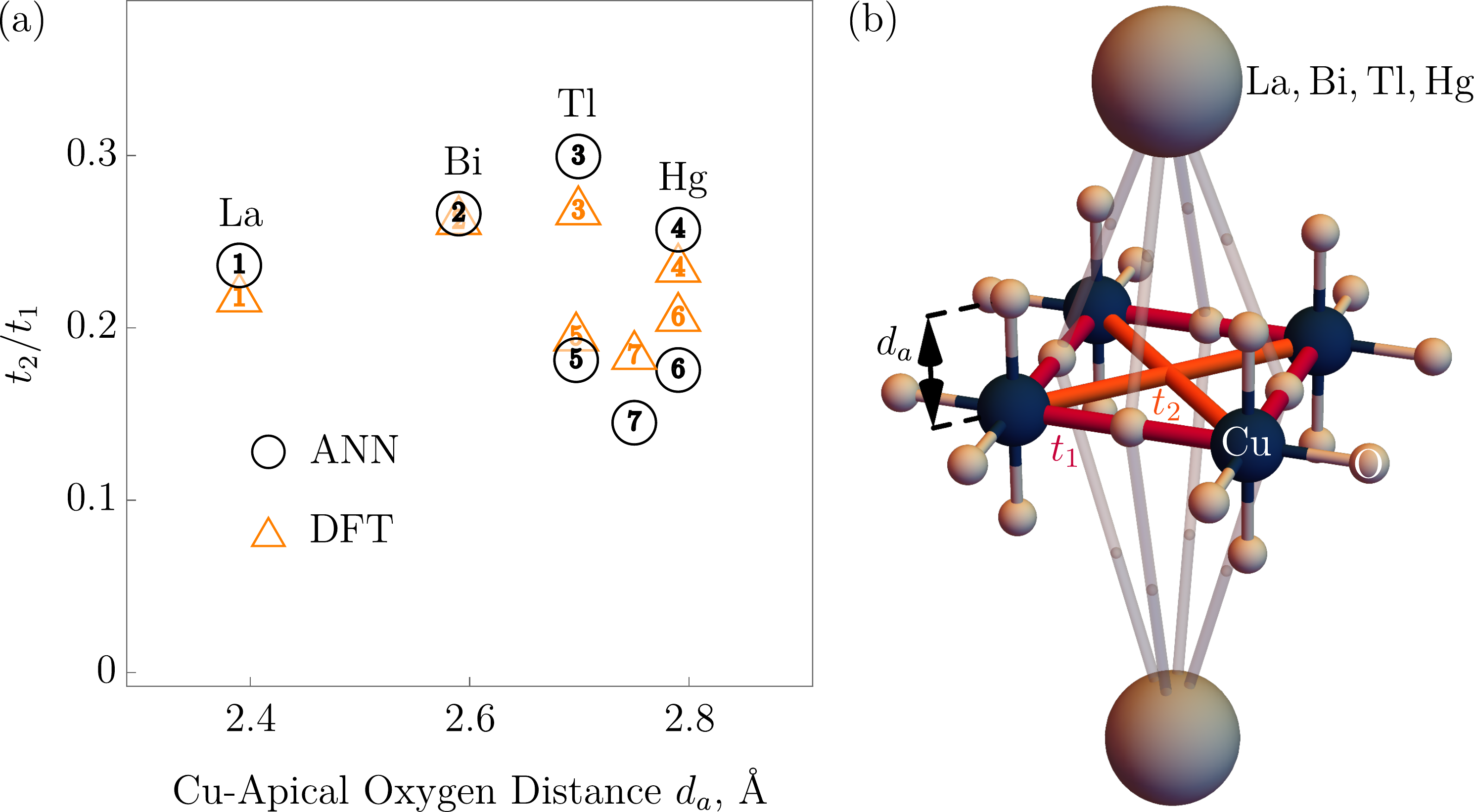}
        \caption{(Color online) (a) The $t_2 / t_1$ ratio as a function of Cu - apical oxygen distance $d_a$ for HTSC cuprates. Black circles corresponds to the predictions of ensemble ANN model, and orange triangles -- results of the DFT calculations.
        Numbers correspond to the following compounds: 1-Tl$_2$Ba$_2$Ca$_2$Cu$_3$O$_{10}$, 2-La$_2$CuO$_4$, 3-Bi$_2$Sr$_2$CuO$_6$, 4-Tl$_2$Ba$_2$CuO$_6$, 5-HgBa$_2$CuO$_4$, 6-Tl$_2$Ba$_2$CaCu$_2$O$_8$, 7-HgBa$_2$CaCu$_2$O$_6$, 8-HgBa$_2$Ca$_2$Cu$_3$O$_8$. (b) Schematics of the distinct structural element of the HTSC cuprate hosting leading $t_1$ and considerably smaller $t_2$ transfer integrals.
        }
        \label{fig:htsc}
        \end{figure}
    \end{center}

To test the accuracy of our ensemble ANN model, we consider the following
parent HTSC compounds: La$_2$CuO$_4$, Bi$_2$Sr$_2$CuO$_6$, Tl$_2$Ba$_2$CuO$_6$,
HgBa$_2$CuO$_4$, Tl$_2$Ba$_2$CaCu$_2$O$_8$, HgBa$_2$CaCu$_2$O$_6$, and
HgBa$_2$Ca$_2$Cu$_3$O$_8$. It is important to note that only the former four
structures were included in the training dataset.  For all compounds, we
recover the frustrated square lattice model with a dominant $t_1$ and a
considerably smaller $t_2$.  For ease of comparison with
Ref.~\cite{Pavarini2001}, we plot the resulting $t_2/t_1$ ratios as a function
of the distance between Cu and the apical oxygen atom ($d_\text{a}$) in
Fig.~\ref{fig:htsc}(a). In the same plot, we show the results of direct
calculations of these transfer integrals by DFT calculations and
Wannierization. A very good agreement is found for all seven cases.

The closely related family of double-perovskite cuprates A$_2$CuTO$_6$ (A = Ba
or Sr, T = Te or W) represents a more challenging test case. Here, the
$t_2/t_1$ ratio crucially depends on the nature of the T atom: in the two
Te-containing compounds, the leading coupling follows the shortest connections
($t_1$), while in the other two compounds the empty $5d$ shell of W boosts
the diagonal coupling ($t_2$)~\cite{katukuri20}. The sensitivity of our ANN
model does not suffice to fully account for this trend: it yields a
dominant $t_2$ for all four compounds.  Despite this shortcoming, the model
correctly reproduces the $t_1$-$t_2$ model, and the predicted $t_2/t_1$ ratio
is lower for Te-containing (1.55 for Sr$_2$CuTeO$_6$ and 1.3 for
Ba$_2$CuTeO$_6$) than for W-containing (1.6 for Sr$_2$CuTeO$_6$ and 1.95 for
Ba$_2$CuWO$_6$) compounds.

Next, we turn to quasi-one-dimensional cuprates.  The dominance of $t_1$ is
correctly reproduced for the quasi-one-dimensional
Sr$_2$CuO$_3$~\cite{rosner97}, another compound with corner-sharing connections
between CuO$_4$ squares. Importantly, this structure features one shorter Cu-Cu
connection, which is not accompanied by sizable electron transfer, and our
model correctly captures this aspect: the respective predicted transfer
integral is about 20 times smaller than the leading intra-chain term.  For
another quasi-one-dimensional compound, linarite PbCuSO$_4$(OH)$_2$ featuring
edge-sharing chains, our model correctly recognizes the relevance of first- and
second-neighbor transfer integrals along the chains, and correctly identifies
the leading interchain coupling~\cite{heinze22}. Importantly, in linarite 
like in many other edge-sharing cuprates, the nearest-neighbor exchange is
ferromagnetic. Such exchanges have a more complex nature and can not be
described within the effective one-orbital model, which is at the core of our
approach. However, the presence of an edge-sharing connection essentially
implies the relevance of the respective magnetic exchange, making a dedicated
estimation of the electron transfer unnecessary.

As a less trivial case, we consider two isostructural natural minerals in which
Cu$^{2+}$ atoms form a kagome lattice: kapellasite
$\alpha$-Cu$_3$Zn(OH)$_6$Cl$_2$ and haydeeite $\alpha$-Cu$_3$Mg(OH)$_6$Cl$_2$.
The relevance of the cross-hexagon coupling $t_d$ and the corresponding
magnetic exchange was suggested based on DFT results~\cite{janson08, jeschke13}
and confirmed experimentally~\cite{fak12, boldrin15}. (As a side note, the
$J_d$ exchange is the principal source of frustration in these systems, because
the nearest-neighbor exchange $J_1$ is ferromagnetic.) Here, we consider
crystal structures of kapellasite and haydeeite that were determined by neutron
diffraction; these structures are not in the ICSD and hence were not used for
training.  The ANN model yields nearly identical results for both materials,
suggesting the leading $t_1\simeq78$\,meV, plus sizable $t_2$ and $t_d$ of
about 25\,meV each. The $t_1$ and $t_d$ values are comparable with the
first-principles calculations~\cite{janson08}. Given that the latter used a
different functional and a different structural input, the agreement is very
good. Yet, the relevance of $t_2$ in the ensemble ANN model is a spurious
result which is at odds with the DFT calculations and experiments.

In all previous examples except Sr$_2$CuO$_3$, the electronic structure
featured several relevant transfer integrals.  There are many cuprates whose
magnetism is shaped by a single coupling dominating over other terms, but it is
unclear which coupling is dominant. An instructive example is the spin-dimer
compound Cu$_2$TeO$_5$, where magnetic dimers do not coincide with the
structural ones~\cite{das08, ushakov09}. For this material, our ANN model
successfully reproduces the magnitude of the strongest coupling and its
dominance over other terms. Another relevant example is the spin-chain compound
CuSe$_2$O$_5$~\cite{janson2009}, where electron transfer is facilitated by the
$[$Se$_2$O$_5]^{2-}$ anionic group connecting two CuO$_4$ squares that are at
an angle to each other. Also here the ANN model correctly identifies the
leading transfer integral.

Naturally, the predictive power of the model is limited, and in some cases the
desired accuracy is not reached.  For instance, in Bi$_2$CuO$_4$ the leading
coupling operates between the structural chains formed by stacks of CuO$_4$
squares, while the nearest-neighbor coupling within these stacks is three times
weaker~\cite{janson07}. Our ANN model correctly reproduces the leading
coupling, yet predicts that the nearest-neighbor coupling has a similar
strength. While the structure of ANN does not allow us to unequivocally
determine the root cause of this discrepancy, we believe that it stems from
the correlation between the magnitude of the transfer integral and the Cu..Cu
separation. While on the average shorter distances indeed correspond to larger
$\|t\|$, in some materials like Bi$_2$CuO$_4$ it is not the case. 

There are several ways to improve the accuracy of the model. An apparent
solution is to extend the dataset by including structures that are not
represented in the ICSD. Also a revision of materials that were filtered out
due to failed Wannierization can make the dataset bigger. However, such
amendments will lead to incremental, moderate improvement of the predictive
power. Based on our analysis, we conclude that main factor limiting the
accuracy of the model are the crystalline environments descriptors. Making them
more specific to chemical environments, e.g.\ by taking the connectivity of
atoms within a chosen sphere into account or a more explicit consideration of
charge densities, and keeping them as compact as possible may significantly
improve the performance of the model. 

To finalize the discussion, we emphasize that the main strength our model is
its ability to identify magnetically relevant couplings. This is particularly
useful for involved structures with a large number of short- and middle-range
Cu..Cu separations, where the leading electron transfer paths can be highly
nontrivial.  The performance remains good across different classes of cuprates,
which allows for efficient screening: evaluation of transfer integrals for a single
material takes between dozens of seconds and a few minutes. Naturally, our
model can not serve as a complete replacement to full-blown first-principle calculations, because
error bars for the individual terms may be too high for certain quantitative analyses. 
However, the model's predictive power is enough to perform qualitative assessment of interactions in spin models.
Furthermore, the developed model holds substantial promise for enabling the inverse construction of hypothetical materials with prescribed magnetic topologies.
For instance, one can create a Cu-O+X network and manipulate its structure to achieve a particular magnetic coupling 
arrangement, as indicated by the transfer integrals data generated by the model. Alternatively, one can begin with a known material 
and inquire about the alterations needed to activate or deactivate, as well as strengthen or weaken, specific magnetic connections.
This then opens a plethora of questions on how these enhanced properties may be received in an actual chemical solid-state structure.
Thereby this method may offer a novel avenue to engineer materials with distinct magnetic properties and unlock these applications, for instance in the fields of magnetic cooling or data storage.

\section{Conclusions}
We constructed an ensemble deep learning model that estimates the magnitude of
transfer integrals in undoped cuprates. These terms underlie the leading
mechanism of the magnetic exchange, and their knowledge is crucial for
correctly determining the microscopic magnetic model. We employed a mapping
onto a three-dimensional Zernike descriptor to describe crystalline
environments that correspond to individual transfer integrals. The resulting
ANN model trained on our high-throughput DFT calculations results can predict
transfer integrals with reasonable error MAE = 18 meV. The model efficiently
differentiates between weak and sizable transfer integrals, which is most
important for estimating the relevant spin model. We discuss the limitations of
this approach and outline ways of improving the numerical accuracy.

\bmhead{Acknowledgments}
We thank Markus Wallerberger, Roman Rezaev, and Dmitry Chernyavsky for fruitful
discussions. This project was funded by the Leibniz Association through the
Leibniz Competition.  Authors acknowledge financial support from the DFG
through the Collaborative Research Center SFB 1143 (Project-Id247310070). We
thank Ulrike Nitzsche for the technical assistance. 

\section*{Methods}

\bmhead{DFT calculations}
    For high-throughput DFT calculations, we used the crystal structures of cuprates downloaded from the ICSD~\cite{bergerhoff.crystallographic.1987} employing the application programming interface (API).
    We performed nonmagnetic DFT calculations with the full potential local orbital code FPLO of version 18.00-52~\cite{koepernik.prb.1999}. Electron exchange-correlation interactions were described using the Perdew-Burke-Ernzerof (PBE) GGA functional~\cite{perdew.prl.1996}. The electron density was converged to within 10$^{-6}$. The reciprocal space mesh was calculated for each structure, accounting for the size of the reciprocal cell~\cite{suppl}.

    One shot calculation of Hellmann–Feynman forces was performed for each cuprate structure. 
    We performed the relaxation of structures containing hydrogen by optimizing atomic coordinates of all H atoms with respect to the GGA total energy. All symmetries of the respective space group were kept during the optimization.
    The Wannier fit of the band structure was performed only for those structures where calculated forces are below 0.1\,eV/\r{A}~\cite{suppl}.

    The Python library pymatgen~\cite{Ong2013pymatgen} has been used broadly in the high-throughput pipeline code to achieve complete automation.

    \bmhead{Machine learning model}
    In the work we consider regression problem $t_{ij} = f(\vec{D})$ which we handle 
    with ML methods.
    For selection of the predictive model we consider shuffle-split and $k$-fold CV. The CV procedures were implemented with scikit-learn Python library~\cite{scikit-learn}.
    The best performance was shown by the ensemble ANN model. We implemented ANN using the Keras API~\cite{chollet2015keras} written in Python for learning platform Tensorflow~\cite{tensorflow2015-whitepaper}. 
    The Supplementary Information 1~\cite{suppl} provides the details on the architecture and training process of the ensemble ANN.





\end{document}